# Live-action Virtual Reality Games

Luis Valente[1], Esteban Clua[1], Alexandre Ribeiro Silva[2], Bruno Feijó[3]

[1]MediaLab, Institute of Computing, UFF, Brazil

[2]Instituto Federal do Triângulo Mineiro, Brazil

[3]VisionLab, Department of Informatics, PUC-Rio, Brazil

{lvalente,esteban}@ic.uff.br, alexandre@iftm.edu.br, bfeijo@inf.puc-rio.br

**Abstract.** This paper proposes the concept of "live-action virtual reality games" as a new genre of digital games based on an innovative combination of live-action, mixed-reality, context-awareness, and interaction paradigms that comprise tangible objects, context-aware input devices, and embedded/embodied interactions. Live-action virtual reality games are "live-action games" because a player physically acts out (using his/her real body and senses) his/her "avatar" (his/her virtual representation) in the *game stage* – the mixed-reality environment where the game happens. The *game stage* is a kind of "augmented virtuality" – a mixed-reality where the virtual world is augmented with real-world information. In live-action virtual reality games, players wear HMD devices and see a virtual world that is constructed using the physical world architecture as the basic geometry and context information. Physical objects that reside in the physical world are also mapped to virtual elements. Live-action virtual reality games keeps the virtual and real-worlds superimposed, requiring players to physically move in the environment and to use different interaction paradigms (such as tangible and embodied interaction) to complete game activities. This setup enables the players to touch physical architectural elements (such as walls) and other objects, "feeling" the game stage. Players have free movement and may interact with physical objects placed in the *game stage*, implicitly and explicitly. Live-action virtual reality games differ from similar game concepts because they sense and use contextual information to create unpredictable game experiences, giving rise to emergent gameplay.

## 1 Introduction

In the past, virtual reality hardware (such as head-mounted display devices) has been restricted mostly to research labs due to factors such as their high cost, the inconvenience of using this hardware (*e.g.,* heavy helmets, lots of cables required, limited mobility), and application specificity (*e.g.,* high end VR systems, military applications). One of the main goals in virtual reality applications is to immerse the user's senses in an artificial virtual environment (VE) through an interactive experience. A key factor regarding how this interactive immersive experience is successful refers to the sense of presence [1]. Recently, there is a growing trend in the industry to bring these kinds of devices to the mass market (*e.g.,* Oculus Rift, Samsung VR, HTC Vive), with affordable prices and small form factors, which opens up possibilities for using these devices in mainstream games and entertainment applications.

The advancement of virtual-reality technologies and context-awareness give rise to new possibilities of game genres. In this regard, in this paper we present a new game genre based on unique combinations of live-action, mixed-reality, context-awareness, and interaction paradigms comprising tangible objects, context-aware input devices, embedded interactions, and embodied interactions. We call this new genre as "live-action virtual reality games".

We say that live-action virtual reality games are "live-action games" because a player physically acts out his/her "avatar" (his/her virtual representation) in the *game stage* – the mixed-reality environment where the game happens. The game creates the *game stage* based on several real-world information sources, such as the physical environment architecture and context information. Live-action virtual reality games keep the virtual and real-worlds superimposed as the player experiences the game. The player sees a virtual world through a HMD device (in first-person perspective) and needs to move and walk in the physical environment to play the game. Also, the player may need to use gestures and other body movements in interactions with game elements and other players. The player may need to interact with physical objects to complete game tasks. Live-action play in live-action virtual reality games is facilitated through light-weight and mobile HMDs, mobile computing, and context-awareness.

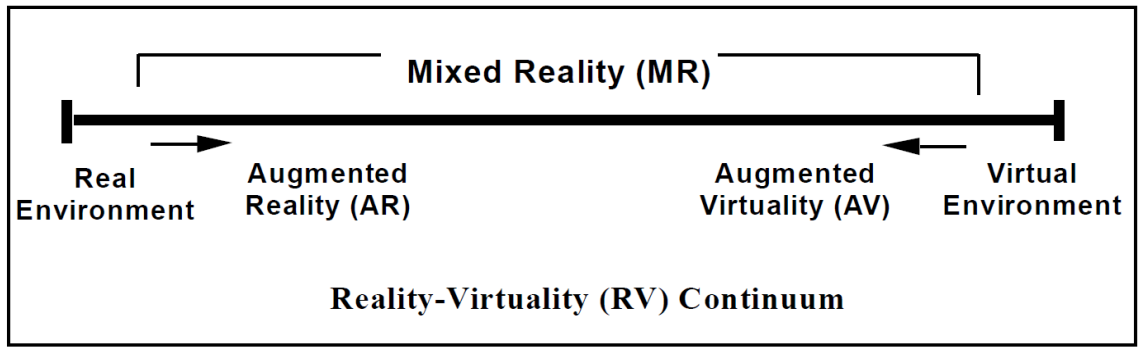

Figure 1. Simplified reality-virtuality continuum, extracted from [2]

This paper is organized as follows. Section 2 presents the live-action virtual reality game vision and its main concepts. Section 3 provides related work, which compares live-action virtual reality games to other types of games based on mixed-realities and initiatives originated in industry and academia. Section 4 presents concluding remarks.

## 2 Live-action virtual reality games

This section presents key concepts of live-action virtual reality games, which are: mixed-reality and the game stage, information flow, context-awareness, and interaction paradigms that live-action virtual reality games provide. This section also discusses how a live-action virtual reality game may stimulate the sense of presence.

### 2.1 Mixed-reality

Live-action virtual reality games create a mixed-reality environment, which we name as *the game stage*. A "mixed-reality environment" can be defined as an environment that fuses virtual and real-world (physical) information. Figure 1 illustrates a simplified "reality-virtuality" continuum that Milgram and co-authors [2] proposed. This continuum represents a spectrum describing possible combinations of virtual and real elements, resulting in different kinds of "mixed-realities".

The *game stage* is a kind of "augmented virtuality" because the virtual world is constructed through real-world information sources and enriched through virtual and real-world information. We roughly categorize these real-world information sources as "physical structure information" and "context information". Live-action virtual reality games output information in the virtual and physical worlds. Physical world output can take form as real-world effects, such as manipulating environment properties (*e.g.*, temperature), generating vibrations, and outputting smells.

### 2.2 Physical structure as input

A live-action virtual reality game uses physical structure information as basic building blocks to create the game stage. When a player enters the physical environment where the game happens (*e.g.*, a room), the game tracks all architectural elements (such as walls) to use as the basic geometry to create a virtual 3D environment. After creating the virtual world structure, the game may augment this raw structure with virtual content. Metaphorically, we compare this process to the process of mapping images (textures) to raw polygon data in 3D environments. The game keeps the virtual and real worlds superimposed as the player moves in the environment, creating the game stage.

The game may also track physical objects that are part of this physical environment and map them in the virtual world as 3D models. Examples of these physical objects are furniture, interactive objects carried by a player, and players' own bodies. However, the player does not see the physical world and physical objects – the player only sees virtual representations through the HMD. The virtual world building process may happen in real-time or before the game session (as a preparation step). In this preparation step, the game system may track all relevant physical features to generate a virtual world to be optimized later by an artist, who may augment the virtual world with other virtual elements.

### 2.3 Context information as input

Context-awareness is a key component of live-action virtual reality games. Dey [3] defines context as *"any information that can be used to characterize the situation of an entity. An entity is a person, place, or object that is considered relevant to the interaction between a player and an application,*

*including the user and applications themselves"*. A system can be considered as "context-aware" if *"it uses context to provide relevant information and/or services to the user, where relevancy depends on the user's task"* [3]. In this regard, "context-awareness" means that a live-action virtual reality game is able to adapt the gameplay according to the current context conditions. Context-awareness enables live-action virtual reality games to collect information from the real-world and from players (while the game session is happening), using these data as input information. Live-action virtual reality games use context information for two main purposes:

1. Augment the game environment with dynamic information that originates in the real-world;

2. Generate game content dynamically to create unpredictable game experiences, giving rise to emergent gameplay.

Here are some examples of information that we consider as "context" in live-action virtual reality games (this list is by no means an exhaustive list):

- Physical environment information. Examples include temperature, humidity, and lighting conditions, time of the day, weather information, and environmental noise;

- Existing research on Human-computer interaction [4] suggests that the concept of context can be boarder enough to consider the ways a user interacts with a device as context information. In this regard, we include as context information about interactions with input devices and with the physical environment where the game takes place;

- Player information. Examples include physiological state, personal preferences, social network profiles, and personality traits;

- Information derived from the possible forms of relationships and interactions among players – the social context [5].

A live-action virtual reality game is able to sense context information through several means. Some possible alternatives are:

1. The player carries or wears devices that sense context from the environment and/or from the player (*e.g.,* physiological information);

2. The physical place where the game happens may host customized devices that are equipped with sensors – the *smart objects*. Smart objects may be connected to other smart objects and to central game servers. These devices may provide interaction interfaces to the players (Section 2.5) and may be able to output information (such as audio and images) and create real-world effects (Section 2.4).

3. The physical place may be augmented with dedicated infrastructure that enables the game to sense context information. Examples include installing tracking technology and sensors to collect environmental information. In this solution, these objects are hidden ("invisible") from players, being part of the overall game infrastructure. On the other hand, *smart objects* are complex, visible physical objects that reside inside the *game stage*;

4. The game queries remote information about the player based on the player identity (*e.g.,* social network profiles).

## 2.4 Information output and real-world effects

Live-action virtual reality games may output information through wearable devices, mobile devices, and *environment devices*. These devices may output digital media information and create real-world effects (*e.g.,* vibrations, smells, smoke). These devices may be connected to other devices that comprise the game infrastructure, such as input devices, game servers, and other output devices. These devices may also be input devices and context-aware.

The wearable devices are devices worn by a player, which may also work as input devices and capable of sensing context information. Examples include HMDs, isolating or non-isolating headphones, gloves, vests, and clothes. These devices might provide haptics feedback as output, for example.

The mobile devices are devices that a player may carry around, touch, and throw away. These devices may output digital information and may be equipped with actuators, motors, and other devices capable of generating real-world effects.

*Environment devices* are objects that reside in the physical environment and generate effects or behavior in the physical world. Some possibilities for *environment devices* are (this is by no means an exhaustive list):

1. Simple output devices, not limited to digital media (*e.g.,* audio speakers, smell generators, weather effects, heat generators, smoke generators);

2. Mechanical or electrical devices equipped with actuators (*e.g.,* sliding doors, elevators, wind generators);

3. *Smart objects* (Section 2.3) equipped with output hardware. An example is the trash can in Fun Theory [6].

## 2.5 Interaction possibilities

In live-action virtual reality games, players interact with the game and with other players through multiple modalities (*e.g.,* voice, body movements, and gestures), ordinary physical objects, customized input devices, and context-aware devices, supporting tangible, embedded, embodied, and context-aware interaction paradigms.

The player may interact with the game by manipulating mobile devices equipped with sensors and ordinary physical objects (*e.g.,* wood sticks, rocks). The mobile devices may be equipped with networking capabilities. These objects and devices realize the "tangible object metaphor", which stands for *"human-machine interfaces that couple digital information to physical objects"* [7]. These interfaces can be implemented through tracking technologies or sensors. Examples of suitable sensors for this purpose are RFIDs, magnetic sensors, proximity sensors, and light sensors. For example, on a general context tangible objects could be pieces of augmented table-top games [8]. Another example is using wood sticks (an ordinary object) that the game tracks to represent swords (the virtual representation in a game). Tangible objects may also have embedded computing capabilities. Existing research suggests that tangible objects contribute to increasing player immersion in games [9–11]. In live-action virtual reality games, players may touch, manipulate, and throw these objects and devices (among other possibilities) to interact with the game. The player does not see these objects and devices directly – the game presents their virtual representation to the player through the HMD. These objects and devices may be carried by the player or may be spread in the game stage. For example, a game may present a puzzle for the player to solve that requires finding an object in the environment and placing it on top of another object or device that resides the environment.

Live-action virtual reality games may use customized input devices or general purpose controllers as interaction interfaces. These devices may include joysticks and custom hardware that features a number of affordances that are specific for the game. An example of customized controller is a controller modeled as gun. An example of general purpose general controller is the Razer Hydra [12]

Another possibility for interaction in live-action virtual reality games is through wearable devices. These devices are devices worn by a player, providing embedded, embodied, and context-aware interaction. Examples of these devices include smart bands, motion capture devices, and brain-machine interfaces. Wearable devices usually are context-aware and may be able to sense the player's physiological state to use as contextual input information.

Players in live-action virtual reality games have to physically walk to move in the mixed-reality environment to complete game tasks. In this regard, players act out their own avatar, which is a form of full body interaction. Full body interactions in live-action virtual reality games can be mediated by wearable devices or not. With the expression "full body interactions not mediated by wearable devices", we mean refer to interaction paradigms made possible by devices such as the Kinect sensor [13].

Another way to interact in live-action virtual reality games takes form as *smart objects* (Section 2.3). *Smart objects* enable interactions of "implicit nature" [14]. In implicit interactions, the interaction interface is invisible (*i.e.,* based on sensors). Schmidt [15] defines an implicit interaction as *"an action, performed by the user that is not primarily aimed to interact with a computerized system but which such a system understands as input"*. For example, a game stage may host a *smart object* containing a proximity sensor that opens a door when someone gets close to it. Implicit interactions occur inadvertently from the player point of view. On the other hand, explicit interactions occur through direct (conscious) player commands, meaning that the player has conscious intention to start something to achieve a goal. As the player does not see the physical world, live-action virtual reality game designers should be careful when designing implicit interactions to avoid accidents that might injure players.

## 2.6 Mixed-reality infrastructure and management

A *game stage* requires a dedicated physical place to create a customized physical installation, due to the infrastructure required to support the mixed-reality in live-action virtual reality games. Before game sessions take place, the required infrastructure is deployed to the physical environment, as a preparation step. Live-action virtual reality games designers and developers may exploit this requirement to their advantage by deploying custom hardware (such as *smart objects*) in the environment that otherwise would not be generally accessible to end users, which might help in creating more sophisticated game experiences. Using a dedicated, customized installation also helps to implement an "uncertainty handling policy" [16] to remove (or minimize) problems caused by limitations of the in-volved technologies, such as networking and sensors.

In live-action virtual reality games, the problems of "tracking physical world features" and "keeping the virtual world properly overlaid on the physical world (*e.g.,* synchronized)" are of central importance. Traditional computer vision techniques (such as the ones based no QR codes) do not have the required accuracy address these problems. So, live-action virtual reality games usually require other solutions based on other kinds of sensors and tracking techniques.

Live-action virtual reality games may require ongoing supervision while game sessions happen. This supervision, also known as "orchestration" [17] aims to anticipate and/or correct technical issues that may happen during a game session, so that the game experience is not broken due to these problems. Orchestration also can be used to help players who are experiencing difficulties in playing the game.

## 2.7 Presence and sense stimulation

Presence can be commonly defined as *"a sense of being in a VE rather than the place in which the participant's body is actually located"* [1]. Important aspects of presence include the extend of the user field of view, number of sensory systems stimulated by the VR system, the quality of rendering in each sensory modality, the quality of tracking, among other aspects [1]. Sanchez-Vives and Slater [1] refers to "immersion" as *"the technical capability of the system to deliver a surrounding and convincing environment with which the participant can interact"*. Immersion and presence are related concepts, but they are not synonyms. However, the literature on these terms is confusing as work by Schuemie and coauthors [18] demonstrate. In live-action virtual reality games, a player experiences the constructed mixed-reality with at least five senses: sight, touch, hearing, kinesthetic sense (*e.g.*, sense of movement and body awareness), and the vestibular sense (*e.g.,* sense of balance).

For example, we provide few possible alternatives on how a live-action virtual reality game could stimulate senses. Live-action virtual reality games stimulate sight through mobile, lightweight HMDs. Live-action virtual reality games may stimulate touch through wearable devices (capable of providing haptic feedback), the underlying physical structure, and physical objects (*e.g.*, touching physical walls, tables, and holding small objects). Hearing may be stimulated by isolating or non-isolating headphones and *smart objects*. Live-action virtual reality games stimulate the kinesthetic and vestibular senses are by requiring players to move in the physical environment and by using interaction interfaces based on gestures. In a customized game stage, live-action virtual reality game designers may deploy *smart objects* and output devices that stimulate other senses, such as smell and the sense of heat.

## 3 Related work

An important aspect of live-action virtual reality games is the mixed-reality. However, the concept of "mixed-reality" can be very broad as Figure 1 illustrates. In this regard, Section 3.1 compares live-action virtual reality games to other kinds of games based on mixed-reality concepts, such as "pervasive games" and "augmented reality games". Section 3.1 also compares live-action virtual reality games with "alternate reality games", due to the idea that live-action virtual reality games transport the player to a "different, alternate" reality – we believe that the term "alternate reality game" might be misleading at first sight for some readers. Section 3.2 compares live-action virtual reality games with existing initiatives found in the industry and academia.

## 3.1 Games based on mixed-reality and virtual reality

This section presents some earlier virtual reality games, pervasive games, and alternate reality games.

### 3.1.1 Early virtual reality games

The Virtuality Group [19] provides earlier examples (in 1990) of using VR in consumer games, as virtual reality arcade machines. These machines contained a HMD, speakers, microphone, and joysticks. That system was able to track head and joystick movements through a magnetic tracking system. Games released in this platform include *Dactyl Nightmare* [20] and *Grid Busters* [21]. Another attempt at bringing VR-based games into the mainstream came from Nintendo and its Virtual Boy [22] in 1995. Although the system was original, it was a commercial failure. King and Krzywinska [23] suggested that the high price and physical discomfort while playing the game contributed to the demise of this device.

The earlier virtual reality games focused mainly on immersing the user's visual sense in the virtual world. The users interacted in the virtual environment with limited mobility – physically walking in the environment was not possible as the equipment running the simulation was not portable.

### 3.1.2 Augmented reality games

Live-action virtual reality games are fundamentally different from games based on "augmented reality" (left side of Figure 1), because in live-action virtual reality games the player does not see the physical world. In augmented reality games the player sees the physical world (through a HMD or a mobile device camera) and virtual contents that are placed on top of physical world elements. On the other hand, in live-action virtual reality games the player sees a virtual world that is built based on physical world characteristics, including its architectural elements and contextual information. A classic reference on augmented reality can be found in [24].

A classic, early example of augmented-reality game is *ARQuake* [25], an indoor and outdoor game where player wears a backpack containing a laptop, an orientation sensor and a GPS sensor, which allows the player to walk freely in a physical environment. The player uses a see-through HMD that is connected to these devices. The *ARQuake* tracking system is based on GPS, digital compass, and fiducial markers. The game takes place on a physical environment that is modeled as a Quake 3D level. When the game is played in the real-world, *ARQuake* uses this 3D model to superimpose virtual world information into the physical world.

### 3.1.3 Pervasive games

The term "pervasive games" refers to games that are played in the real-world, exploring mobility, mixed-realities, and context-awareness, among other aspects. This description of pervasive games is not a consensus in the literature [26].

This kind of game aims to bring the game experience out of a game device into the real world. According to the Oxford English Dictionary [27], *"pervasive" means: "having the quality or power of pervading; penetrative; permeative; ubiquitous"*. This may suggest that "pervasive games" are games "pervading something" (the real-world perhaps) or spread somewhere. In games, pervasiveness can be recognized every time the boundaries of playing expand from the virtual (or fictional) world to the real world, creating mixed-realities. The literature presents various interpretations and scopes for defining what "pervasive games" mean. These interpretations can be roughly divided into "cultural perspectives" (*e.g.,* game design, game studies, which take "pervasive" in its literal sense) and "technological perspectives" (*e.g.*, ubiquitous and pervasive computing). There are some works in the literature that analyze existing interpretations of pervasive games [26, 28]. Also, in the literature there are other terms that are used as synonyms of "pervasive games", such as "mixed-reality games" and "ubiquitous games".

Through a technology perspective, a pervasive game can be considered as a context-aware system that has these characteristics:

- The game is constantly coming back to the real-world, which means that the game is played in physical places and it is not constrained to stationary computing devices;

- The physical world (places, objects) is part of the game and it is combined with the virtual world, creating a mixed-reality;

- Mixed reality is always existent and it is created through pervasive computing technologies (*e.g.*, sensors, context-aware systems);

- The spatial mobility occurs in a physical "open" environment, that is, the "game world boundary" is not "well-defined", and sometimes it can be unconstrained;

- The players use mobile devices (*e.g.,* smartphones, tablets, custom hardware) to interact with the game and with other players;

- The game may last for several days or weeks (or more), blending with the daily lives of players. The game may define a persistent world that progresses without player intervention. If some important event happens in the game, the game may notify the player to take action. These aspects are not mandatory;

- The game may focus on focuses on promoting social interaction among players. Social interaction in a pervasive games may happen directly (face to face interaction) or indirectly (mediated through technology). This aspect is not mandatory as pervasive games may be single-player games.

As live-action virtual reality games, pervasive games also define a mixed-reality. Pervasive games may also use the elements that we described in Sections 2.2 to 2.6. However, the mixed-reality in pervasive games is of different nature than the one found in live-action virtual reality games. Pervasive games are based on the idea of a real-world augmented with virtual content. This augmentation takes form as using context information collected through sensors placed in the physical environment or in mobile devices that the players use. This augmentation may also take form as "augmented reality" (Section 3.1.2). The reader can be referred to [29] for a report describing some pervasive games.

### 3.1.4 Alternate reality games

The term "alternate reality game" (ARG) may suggest that these games bring the player to a kind of reality that is very different from a real-world setting. This idea is metaphorically similar to some of the ideas in live-action virtual reality games, but these game styles are very different. ARGs suggest a surrealistic setting where the game denies its existence "as a game". The main slogan of those games is "this is not a game". ARGs use the real-world as a platform and create a comprehensive interactive narrative, as massive puzzles that span on different media such as web-sites, emails, and phone calls. Game masters create (real-world and virtual) content and steer the story according to players' reactions. The game is purposely ambiguous, so that players always question if the game activities are indeed part of the game, or part of real-world life. This includes discovering how to enter the game and guessing if it is over.

An example of early ARG is *The Beast* [30], which was part of a marketing campaign for the movie *A.I. Artificial Intelligence* [31]. The game began with a question, "Who killed Evan Chan", and then evolved to an interactive story that had been deployed over the internet and the real world. The game itself was not advertised as a game, and its entry-point was hidden in A.I. movie trailers and posters. After following the clues, the player could access "real-world elements" (like voice mails from the game) that opened-up the gate for the storyline. The game designers have created fake websites and other multimedia content to support the game through puzzles and other interactions. Also, sometimes the game would make phone calls to the players. Some researchers classify ARGs as a kind of "pervasive game" [32, 33]. The reader should refer to [34] for more information about ARGs.

## 3.2 Related initiatives

This section presents some industry and academic initiatives related to live-action virtual reality games, ending with some comparisons among them.

### 3.2.1 Industry initiatives with similar concepts

There are some industry and academic initiatives that have similarities with the concept of live-action virtual reality games that we present in this paper. These initiatives seem to be "works in progress" as information about their current status (especially concerning technical details) is scarce. Also, it seems that these initiatives have not debuted yet.

The first on is "The VOID" [35], which is a project envisioned by Ken Bretschneider that is planned to launch in 2016. The VOID defines a mixed-reality based on the real world structure, where the players have to walk to move in the mixed-reality. The environments where the game happens are named as "gaming pods". They intend to provide different kinds of sensations for players, including *"elevation changes, touching structures & objects, feeling vibrations, air pressure, cold & heat, moisture, simulated liquids and smell"* [36]. Players will use a proprietary HMD and wear proprietary vests and gloves, both capable of providing haptics feedback. The players do not see the physical world.

Another related initiative comes from Kenzan [37], who defined their vision as "real virtuality". In the vision by Kenzan, players wear HMDs and have their body and movements captured motion capture devices installed in the physical environment installation. The players also have to walk physically to move in the mixed-reality environment. It seems that players may also interact with physical objects inside the environment, while the game displays their virtual representation (similar to the idea of interacting with physical objects described in this paper).

Survios' ActiveVR [38] is another related vision, which is based on six elements: immersion (*"the perception that a digital environment really exists"*), presence (*"the perception that you exist within a digital environment"*), embodiment (*"the perception that you are physically interacting with a digital environment"*), free movement (*"being completely unbound from physical restrictions in the outside world"*), shared space (*"the perception that you are naturally communicating with other people in a digital environment"*), and dynamic spectating (*"being able to view and interact with people playing in VR from the outside world"*). At the time of writing this paper, Survios's website does not provide more information on how they implement this vision. Survios seems to have originated from "Project Holodeck" [39]. The game demo *Zombies on the Holodeck* [40] from Project Holodeck features users wearing a Oculus Rift device and interacting with the environment using wired Razer Hydra controllers. The demo uses these controllers to create a tangible interface. The Razer Hydra controllers are able to detect motion and orientation through embedded sensors.

VRcade [41] envisions a system with *"immersive head mounted displays, motion capture systems, and real-time 3D engines"* that takes place in *"out-of-home installations as small as 15'x15' or as large as 150'x75' (~11,000 square feet), users are able to explore and interact with virtual envi-*

*ronments with no wires and completely intuitive movements"*. At the time of writing this paper, VR-cade's website does not provide more details about their vision and how their system works.

### 3.2.2 The "Holodeck" vision and related initiatives

StarTrek's Holodeck [42] seems to be an idealized and prototypical model of immersive virtual environment for many virtual reality researchers. The Holodeck corresponds to an empty room that is filled with real content that is generated according to a specific program, creating a "simulated reality". The Holodeck is described in StarTrek's website as:

> "The generic name, especially in use aboard Federation starships, for the "smart" virtual reality system as evolved by the 2360s — a technology that combines transporter, replicator, and holographic systems.
>
> The programs, projected via emitters within a specially outfitted but otherwise empty room, can create both "solid" props and characters as well as holographic background to evoke any vista, any scenario, and any personality — all based on whatever real or fictional parameters are programmed."
>
> While personal holoprograms relieve the stress and isolation of shipboard life for crew personnel, Holodecks are also used for tasks ranging from scientific simulation to tactical or even covert training. Off starships, many commercial users have equipped facilities with so-called Holosuites.

Inspired by this original idea, the "NYU Holodeck" is an initiative from the MAGNET group [http://magnet.nyu.edu] at New York University to create a virtual reality environment with mobile HMDs. There is little information about this initiative so far – the few ones take form as YouTube videos such as [43].

Another unnamed (as far as we know) initiative to create a "Holodeck-room" comes from the Max Planck Institute for Biological Cybernetics. This VR room takes form as the TrackingLab [44], which is a (12.7 m x 11.9 m x 6.9 m) room equipped with 16 motion capture cameras. The user wears an Oculus Rift device equipped with trackers that are captured by the cameras in the room. Joachim Tesch provides a demonstration of this system in a video interview [45].

Marks and co-authors [46] describe yet another concept based on StarTrek's Holodeck, also named "Holodeck". The Holodeck described by Marks and co-authors [46] comprises a physical tracking room, VR headsets (Oculus Rift), and render engines. The tracking room corresponds to a (6 m x 6 m x 3 m) room with 24 cameras to create a motion capture system based on optical markers. In this Holodeck, a user is able to interact with hand-based gestures. The focus of this initiative is visualization of scientific and engineering data.

### 3.2.3 Comparing live-action virtual reality games with the initiatives formerly described

As far as we know, the main difference among live-action virtual reality games and the initiatives described formerly is the augmentation of a virtual world through context-awareness (especially through smart objects). In live-action virtual reality games, we also support the idea of implicit interactions through smart objects.

The key issue being worked in these initiatives relates to keeping the virtual and physical worlds correctly superimposed (or "tracked") so that a user is able to touch physical environment elements (such as wall) and feel like he/she is touching the virtual wall that he/she sees.

We have not been able to find much information about the current state of the industry initiatives that we cited in this paper. The one that seems to be most advanced up to now is The VOID.

As far as we were able to know, the initiatives inspired by the Holodeck vision seem to be in an informal state when it comes to define the core concepts that they want to support. The information we were able to get so far about these initiatives illustrate that in the current state, these systems focused on getting a user to walk in a physical environment while displaying a virtual world. The key issues being worked relate to tracking technologies.

## 4 Concluding remarks

There is a growing trend in the industry to bring VR hardware with affordable prices to the mass market. This trend has reignited VR research. Affordable, lightweight, mobile, and powerful VR hardware facilitate new possibilities such as live-action virtual reality games.

In this paper we outlined a vision about a new genre of digital games – live-action virtual reality games (live-action virtual reality games). Key concepts in live-action virtual reality games are live-action, mixed-reality, context-awareness, and interaction paradigms based on tangible objects, embedded, embodied, and context-aware interactions.

In live-action virtual reality games, players enter a game stage where they interact with the game and other players through several devices and physical objects. The game requires the player to act out his/her avatar in game, which requires him/her to walk, move, interact with physical objects, and use different interaction modalities. The game stage is mixed-reality environment created according to the physical structure where the game stage resides. The game stage is augmented with environment devices (physical objects placed in the physical environment) and context information. Smart objects enable live-action virtual reality games to sense information about the physical environment and the current situation of players, which may open up possibilities to create unpredictable game experiences and emergent gameplay.

There are some research and industry initiatives trying to come up with concepts that are similar to live-action virtual reality games. We were not able to find detailed information about these visions and their current status, especially in the industry side. It seems that many of them are focusing in solving the tracking problem related to keeping the virtual and real worlds correctly superimposed, so that a user is able to walk in both environments simultaneously. This also enables the user to touch physical walls and "feel" their virtual representation, which surely contributes to increase the sense of presence in the virtual environment. As far as we know, context-awareness is a key aspect that differentiates live-action virtual reality games from other related initiatives.

## Acknowledgments

The authors would like to thank CNPq, CAPES, FINEP, NVIDIA, and FAPERJ for the financial support to this research project.